\documentclass[aps,twocolumn,pra,showpacs]{revtex4}
\usepackage[dvips]{graphicx}
\usepackage{dcolumn}
\usepackage{bm}
\begin{document}

\author{Jakub S. Prauzner-Bechcicki$^1$, Krzysztof Sacha$^1$, 
Bruno Eckhardt$^2$, and Jakub Zakrzewski$^1$}
\affiliation{$^1$Instytut Fizyki Mariana Smoluchowskiego and
Mark Kac Complex Systems Research Center \\
  Uniwersytet Jagiello\'nski, Reymonta 4, 30-059 Krak\'ow, Poland \\
$^2$Fachbereich Physik, Philipps-Universit\" at Marburg, D-35032
Marburg, Germany}

\title{Time resolved quantum dynamics of double ionization in strong 
laser fields}

\date{\today}

\begin{abstract}
Quantum calculations of a 1+1-dimensional model for double ionization 
in strong laser fields are used to trace the
time evolution from the ground state through ionization and rescattering
to the two electron escape. 
The subspace of symmetric escape, a prime characteristic
of nonsequential double ionization, remains accessible by 
a judicious choice of 1-d coordinates for the electrons.
The time resolved ionization fluxes show the onset of single and
double ionization, the sequence of events during the pulse, and the
influences of pulse duration, and reveal the relative 
importance of sequential and non-sequential double ionization, even
when ionization takes place during the same field cycle.
\end{abstract}
\pacs{32.80.Rm,32.80.Fb,03.65.-w,02.60.Cb,02.60.-x}

\maketitle

Double ionization in intense laser fields has been challenging because
of a yield much higher than derived from independent electron 
calculations, thus demonstrating the significance of electron
interations (see \cite{review} and references therein). High
resolution experiments revealed that the two outgoing electrons 
preferably leave the atom side by side,
with the same parallel momenta \cite{review,weber00n}. 
The theoretical understanding and interpretation of this process 
is still far from being complete.
The most accurate representations of the process, i.e.
the exact solution of the time-dependent 
Schr\"ordinger equation for two electrons in a laser field \cite{taylor}
or S-matrix calculations \cite{becker00kopold00} are computationally
demanding and still do not fully represent the experiments. 
Low-dimensional models frequently sacrifice the experimentally
dominant subspace of symmetric escape by restricting the
electrons to move along a common line (aligned-electron models) 
\cite{alinged,engel}, or introduce other correlations, as in \cite{becker},
where the motion of the electron center of mass is restricted to be 
along the field polarization axis. 
The 1+1-dimensional model we present here removes these drawbacks, 
and allows for efficient calculations which give 
time- and momentum-resolved insights into the dynamics of 
the process, from the turn on of the field to the final 
escape of the electrons. The calculations reveal the time order of
events and clearly show the significance of the saddle in the
symmetric subspace for the ionization yield, thus lending further support to the 
ideas derived from studies of the classical models \cite{ours}.

The model is motivated by the rescattering scenario,
\cite{corkum93}. While most electrons
leave the atom directly and contribute to the single ionization 
channel, some have their paths reversed by the field and return
to the core. The acceleration by the field brings in enough energy
so that when this energy is shared with another electron close to the 
nucleus, each has enough energy to ionize.
During the collision with the other electron, a short lived 
compound state is formed which then decays into different 
possible channels: double ionization, single ionization or 
a repetition of the rescattering cycle. Starting from this
intermediate state a classical analysis easily
yields possible pathways to ionization 
\cite{ours}. The classical model of non-sequential 
double ionization (NSDI) suggests that the electrons may escape 
simultaneously if they pass sufficiently close to a saddle in
the symmetric subspace that form in the presence of the electric field. 
As the field phase changes, the saddles for this correlated 
electron escape move along lines that keep a constant angle 
with respect to the polarization axis. 

The observation that the saddles move along
lines through the origin suggests a model where
each electron is confined to move along this reaction coordinate 
\cite{eckhardtunpubl}.
The main difference of this model 
with the aligned electron models is
that a symmetric motion of the electrons is possible and 
not blocked by electron repulsion. 
The present model is hence able to reproduce tunneling and 
rescattering processes, single and sequential double ionizations 
and it correctly mimics the correlated electron escape.  
Moreover, because of the restriction
to 1+1 degrees of freedom it can be integrated by standard methods.

Taking into account that the lines form an angle of $\pi/6$ with the 
field axis \cite{ours}, the restricted classical Hamiltonian for the 
two electrons in the linearly polarized laser field is given by 
(in atomic units)
\cite{eckhardtunpubl}:
\begin{equation}\label{ham}
H=\sum_{i=1}^2\left(\frac{p_i^2}{2}-\frac{2}{|r_i|}+\frac{F(t)\sqrt{3}}{2}r_i\right)+
\frac{1}{\sqrt{(r_1-r_2)^2+r_1r_2}},
\end{equation}
where $r_i$ are the electron coordinates along the saddle lines. 
The electric field is $F(t) = F f(t) \sin(\omega t +\phi)$,
with amplitude $F$, envelope $f(t)$, frequency $\omega$ 
(here: $\omega=0.06$~a.u.) and phase $\phi$. 
For a fixed time and field $F(t)$ the potential energy 
(\ref{ham}) has a saddle located in the invariant symmetric 
subspace $r_1=r_2$ and $p_1=p_2$
at $|r_1|=|r_2|=3^{1/4}/\sqrt{|F(t)|}$ of energy $V_s(t)=-3^{3/4}2\sqrt{|F(t)|}$
\cite{footnote}.
If the electrons pass close to this saddle and sufficiently 
close to the symmetric subspace, they can leave the atom simultaneously. 
The saddle defines the bottleneck for simultaneous escape. 
Once the electrons are outside the barrier they are accelerated 
by the field and escape. Any asymmetry of electron motion 
around the saddle can be amplified by the field. Thus, even if 
they escape simultaneously their final momenta can be quite different
(see Ref.~\cite{eckhardtunpubl} for examples of classical
trajectories in a static field).
In this letter we analyze the quantum dynamics of this model by
solving the time-dependent Schr\"odinger equation numerically 
by a Fourier methods.
The Coulomb singularities in the potential in (\ref{ham}) 
are smoothed by the substitution $1/x\rightarrow 1/\sqrt{x^2+e}$ with $e=0.6$,
which leads to a ground state energy of the unperturbed atom of $E_g=-2.83$. 

\begin{figure}
\includegraphics[width=0.5\textwidth]{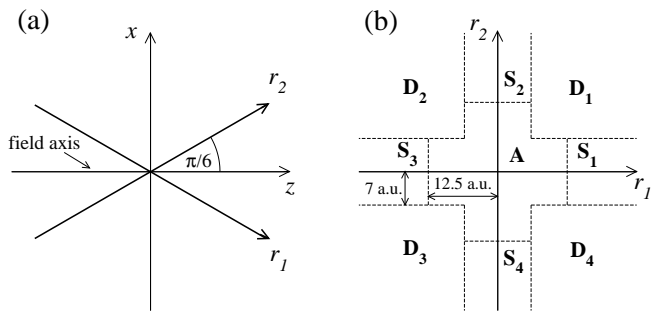}
\caption[]{Geometry of the model. The left frame shows the lines along
which the electrons can move. The right frame shows the division of 
the configuration space $r_1$-$r_2$ into domains assigned to
the neutral atom {\bf A}, singly charged ions {\bf S$_i$}
and doubly charged ions,
{\bf D}$_i$, each with indices $1,\ldots,4$. 
The polarization axis points along $z$.
\label{geometry}}
\end{figure}

For the analysis of the outgoing electrons, we follow Ref.~\cite{taylor} and 
define regions in the configuration space that correspond to the 
neutral atom ({\bf A}), the singly ionized ({\bf S}$_i$) 
and the doubly ionized atom ({\bf D}$_i$) (see Fig.~\ref{geometry}). 
These definitions are suggested
by practical considerations and correspond effectively to a truncation
of the long range effects of the Coulomb potential at these distances.
The regions allow us to distinguish
between the sequential and the non-sequential (simultaneous) 
double ionization by calculating the
probability fluxes between the appropriate regions:
i) The population of the singly ionized states (SI) 
at a time $t$ is obtained from the time integration of the fluxes
from {\bf A} to {\bf S$_i$} minus the fluxes from {\bf S$_i$}
to {\bf D$_j$};
ii) The population of NSDI states is obtained 
from the time integration of the fluxes from {\bf A} to {\bf D$_1$} and 
from {\bf A} to {\bf D$_3$}.
iii) Integration of the fluxes from {\bf S$_i$} to {\bf D$_j$} 
gives a measure of sequential double ionization (SDI) 
processes.
The fluxes from {\bf A} to {\bf D$_2$} and from {\bf A} to {\bf D$_4$}, 
correspond to anti--correlated double ionization: they give negligible 
contributions to the double ionization process and
will not be considered further here. Note that the
definition of the fluxes allows us to distinguish two contributions
to the instantaneous double ionization yield: electrons may pass
directly from region {\bf A} to {\bf D$_1$}, say, or they may
first cross over to {\bf S$_1$} and then to {\bf D$_1$}. The essential difference
is that in the first case electron interactions are significant,
whereas in the second case the electrons remain sufficiently far apart
that their interactions are negligible. After a rescattering event
both processes can happen. This suggests that the definition of 
NSDI should be restricted to processes 
where electron interactions are important, i.e. the direct flux
from {\bf A} to {\bf D$_1$} and {\bf D$_3$}. In the following we will
use the term only in this meaning.

\begin{figure}
\includegraphics[width=0.5\textwidth]{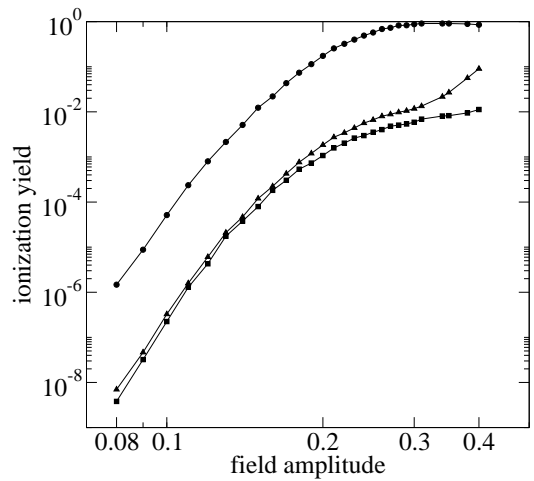}
\caption[]{Yields for single ionization (circles), sequential double (triangles) 
and non-sequential double (squares) ionizations as a function of the field amplitude. The data are obtained for the initial phase of the field 
$\phi=0$ and the pulse envelope $f(t)=\sin^2(\pi t/T)$ where pulse 
duration $T$ equals 5 field cycles. }
\label{yield}
\end{figure}

\begin{figure}
\includegraphics[height=0.35\textwidth,width=0.5\textwidth]{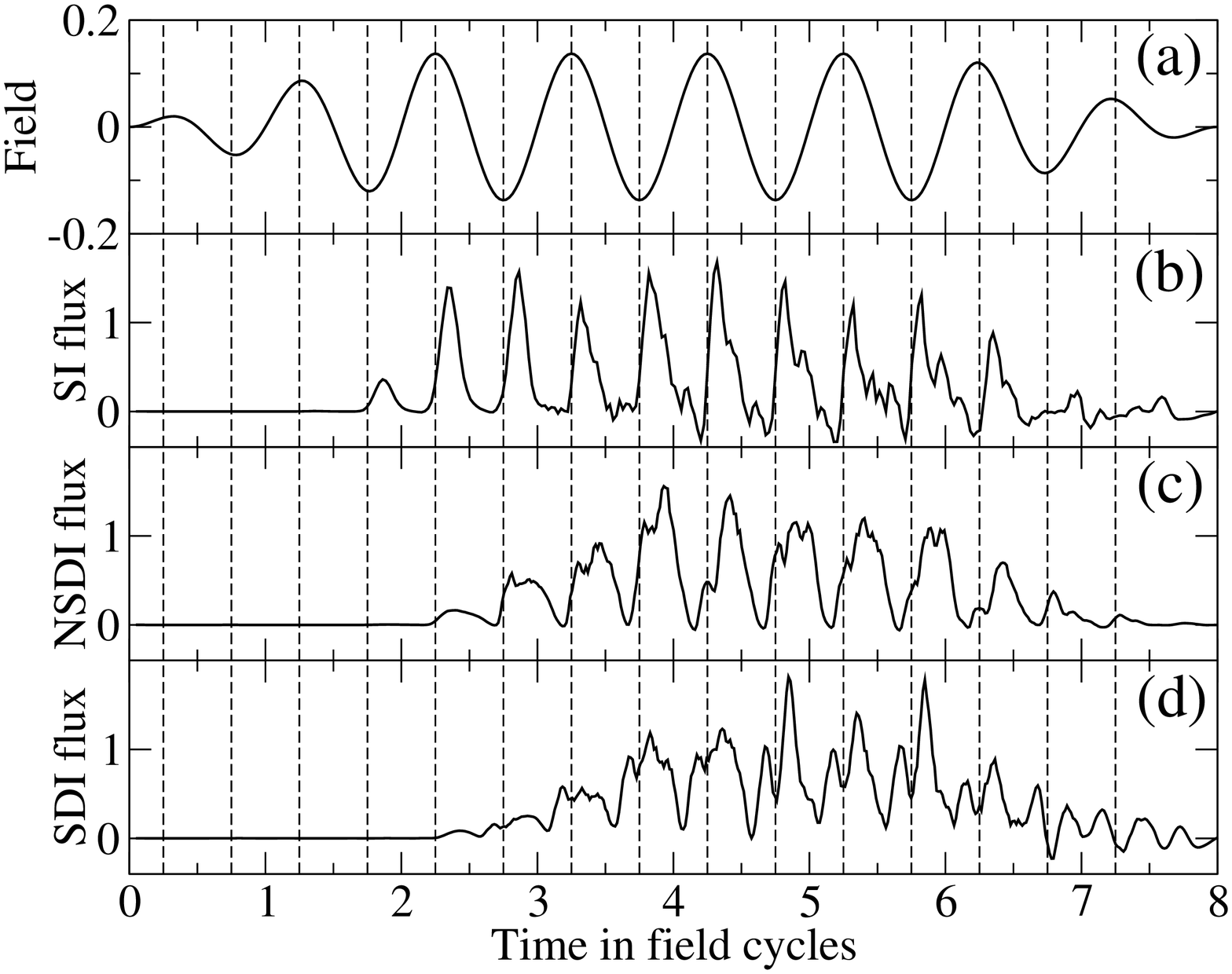}
\caption[]{Probability fluxes (in arbitrary units) as a function of time. 
Panel (a) shows the field strength for $F=0.16$. 
Panel (b) shows the flux related to the single ionization and panels 
(c) and (d) show the non-sequential and sequential double ionization yields, respectively.
The field has an initial phase $\phi=0$, a duration of 
8 cycles, and is switched on and off linearly over 2 cycles. }
\label{time}
\end{figure}

As a first result we show the ionization yields for the different
subspaces in 
Fig.~\ref{yield}.  They are calculated from the
fluxes as described in the preceeding section.
For intermediate fields the NSDI
and SDI signals are about equal, 
but for higher fields SDI rises sharply, forming
the well known knee: in hindsight, it is clear that only the sequential
double ionization can show the knee, as it derives from the strong
increase of independent electron ionization at high fields.
The NSDI-signal, on the other hand, seems to saturate for field
amplitudes above about 0.25 a.u.

The time ordering of the process, resolved into the different 
fluxes, is shown in Fig.~\ref{time} for a field strength of $F=0.16$,
below the knee. Up to the fourth extremum the field is not strong
enough to ionize any electrons. Shortly thereafter
singly charged ions appear, but no doubly charged ones. 
Immediately after the fifth extremum a strong single ionization is 
observed as well as a first double ionization signal. Note that 
the maximum of double ionization occurs after the 
extremal field strength, at about the same time as the 
maximum in the single ionization signal. This shows that in
contrast to expectations based on {\it the simple man model} 
\cite{review,corkum93} for
the returning electron, the double ionization events occur
when the field is still on. 

\begin{figure}
\begin{center}
\includegraphics[height=0.3\textwidth,
width=0.45\textwidth]{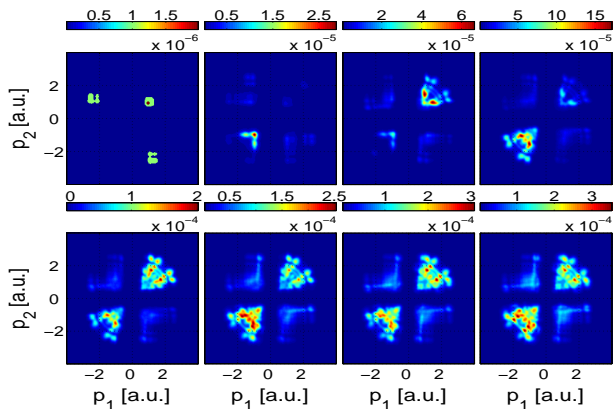}
\end{center}
\caption[]{Time resolved electron momentum distributions 
corresponding to the field parameters in Fig.~\ref{time}.
The times equal the extrema of the field strengths, indicated
by the dashed lines in Fig.~\ref{time}, starting with the one
at $t=3.75$ cycles (top left) and ending with $7.25$ (bottom
right).}
\label{mom1}
\end{figure}

The momentum distributions of the outgoing electrons are obtained following  
the method proposed in \cite{engel} where the wavefunction is propagated
with all interactions in a region of width 400~a.u.$\times$400~a.u.
Outside this domain the wave function is transformed to the momentum space 
where time evolution (with neglected Coulomb potentials and in 
the velocity gauge) becomes simply a multiplication by a time dependent 
phase. Fourier transforms of the parts of the wavefunction 
in the regions $|r_1|,|r_2|>200$~a.u. then give the momentum 
distributions in Fig.~\ref{mom1}. 

The panels in Fig.~\ref{mom1} are calculated for the same
parameters as for Fig.~\ref{time} and give the momentum distribution
at successive extrema of the field. The sequence starts with
the extremum at 3.75 cycles, as there is no noticable wave function
amplitude in the range $|r_1|,|r_2|>200$~a.u. for earlier extrema. 
The wave function that gives rise to the NSDI near times 
of $t=2.25$ oscillates with the field and extends into this space 
region only about 1.5 cycles later. Thus, with the delay taken 
into account, the first signals in the NSDI sector in 
Fig.~\ref{mom1} correspond to the first signals in Fig.~\ref{time}. 
The momentum distributions in the first four panels in 
Fig.~\ref{mom1} start out very much concentrated along the 
diagonal $p_1=p_2$. This confirms that the bottleneck for 
double ionization are the saddle configurations
in the symmetric subspace, described in \cite{ours}. 

After a few cycles, the different ionization signals in 
Fig.~\ref{time} and the momentum distributions in Fig.~\ref{mom1}
experience significant spreading and distortion. They then
no longer reflect the sequence of extrema and rescattering events,
and the temporal relation between the individual processes gets blurred.
Moreover, as time goes on, less correlated and purely sequential
processes become more important. This suggests that the 
structure of the process can best be resolved with short
pulses, say up to 3 field cycles. This is shorter than
the pulses used so far\cite{weber00n}, 
but now within experimental reach \cite{krausz}.

\begin{figure}
\includegraphics[width=0.35\textwidth]{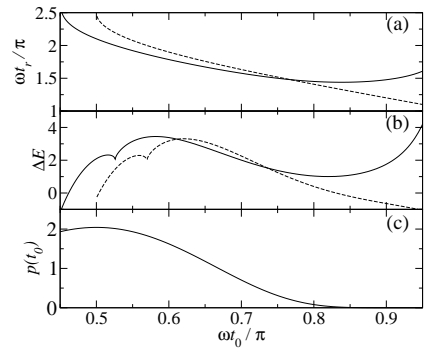}
\caption[]{Rescattering in a 1-d model at $F=0.16$. 
Panel (a) shows the return time
$t_r$, panel (b) the excess energy at the recollision moment 
$t_r$ and panel (c) the tunnelling probability, obtained from a 
semiclassical estimate $\propto e^{-S}$ where $S$ is the action of a 
tunneling trajectory, in unscaled units. The abscissa for all panels
is the point in time where the electrons tunnel through the barried 
(their initial energy is $-0.83$). Dashed lines in panels (a) and (b) 
show the results of the model with neglected Coulomb potential 
\cite{review,corkum93}.}
\label{resc}
\end{figure}

\begin{figure}
\begin{center}
\includegraphics[width=0.40\textwidth]{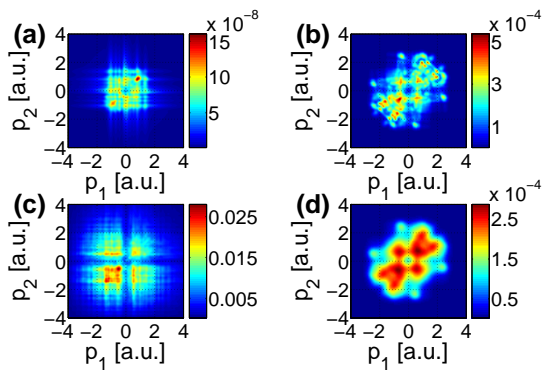}
\end{center}
\caption[]{Final electron momentum distributions for different field
strengths. The pulse is 8 cycles long with a linear switch
on and off over 2 cycles. The momentum distributions are averaged
with a Gaussian of width 0.07 a.u. to model experimental resolution
(and to remove finite size fluctuations from our numerical grid).
The field phase is $\phi=0$ and the amplitudes are $F=0.08$ (a), 
$F=0.16$ (b) and $F=0.4$ (c). 
In order to show the effects of a varying phase and broader smoothing,
panel (d) shows results for $F=0.16$, but averaged over an undetermined
phase $\phi$ and a wider Gaussian window of 0.2 a.u.
The data are the Fourier transforms of the parts of the wavefunction 
(evolved one more cycle after the pulse is gone) 
in the regions $|r_1|,|r_2|>100$~a.u. 
}
\label{mom}
\end{figure}

The time ordering of the process and in particular the
presence of the field when the electrons return to the 
nucleus in the rescattering event, can also be understood 
from the classical dynamics, as in \cite{corkum93}, if the Coulomb field is 
taken into account. To this end we show in Fig.~\ref{resc}
the results from a classical trajectory calculation. 
As in \cite{corkum93} we assume that an electron
that tunnels out is released with zero momentum at the other
side of the potential barrier. It is then integrated classically
until it returns to the atom. Since the process involves
motion of a single electron along the field axis, we can take
a 1-d Hamiltonian,
$H_1=p^2/2-1/\sqrt{r^2+e}+r(\sqrt{3}/2)F\sin(\omega t)$.
The electron that tunnels through the Stark barrier starts with
an energy $-0.83$ a.u., equal to the energy difference between the ground
state of a He atom and a He$^+$ ion in our model\cite{corkum93}. 
If the field is weak,
the electron starts far from the core and can aquire considerable
energy while the field brings it back. However, such processes are very
unlikely since the tunneling probability is negligible.
The relevant energy parameter when the electron returns to the nucleus
is the difference between energy of the two electron system $E(t_r)$ and 
the potential energy of the saddle $V_s(t_r)$ \cite{ours,eckhardtunpubl}, 
defining $\Delta E=E(t_r)-V_s(t_r)$. 
The data collected in Fig.~\ref{resc} (corresponding to $F=0.16$) clearly
show that most electrons return while the field is still on 
in agreement with the sequence of events documented in Fig.~\ref{time}.
For smaller $F$ the range of positive excess energy shrinks 
and moves towards larger values of $t_0$ where the tunneling 
probability is negligible. This indicates
that as $F$ increases one cannot expect a sharp threshold 
behavior for the correlated simultaneous escape since
the contributions from the correlated events grow smoothly with 
field amplitude.

Fig.~\ref{mom} shows results for an 8 cycle pulse and different
field strengths. 
The figure can be compared with 
Fig.~1 of Ref.~\cite{engel} obtained in the aligned electron model 
where due to the overestimated Coulomb repulsion 
the area around $p_1=p_2$ is not populated. Here, this region
is accessible and provides information about the correlated
electron escape. 
With increasing $F$ the double ionization
signal increases, but above the knee ($F=0.4$) the strong contributions
in the second and fourth quadrant show the strong influence of 
SDI. The distributions for undetermined phase
and wide averaging in Fig.~\ref{mom}d still show the strong concentration
of the momentum near the diagonal, but the interference structures
in the corresponding Fig.~\ref{mom}b are washed out.  
The investigation of these interference patterns is 
the subject of ongoing work.

The 1+1-dimensional model for double 
ionization of atoms in strong laser fields discussed here gives 
access to the time-dependence of events and the distribution
of final momenta, including the subspace of symmetric escape.
The good reflection of experimental observations suggests
that the model can also be used to study other aspects of
the ionization process, such as the effects of different
pulse shapes or wave packet interferences.

We thank the Alexander von Humboldt Foundation, 
Deutsche Forschungsgmeinschaft, Marie Curie TOK COCOS scheme, 
and KBN through grants PBZ-MIN-008/P03/2030 (JZ) and 
Polish Government scientific funds (KS:2005-2008, JPB:2005-2006)
 for support.


\begin{thebibliography}{10}

\bibitem{review}
A. Becker, R. D\"orner, and R. Moshammer, 
J. Phys. B {\bf 38}, S753 (2005).

\bibitem{weber00n}
T. Weber {et al.},
Nature {\bf 405}, 658 (2000);
B. Feuerstein {et al.},
Phys. Rev. Lett. {\bf 87}, 043003 (2001);
R. Moshammer {et al.},
J. Phys. B {\bf 36}, L113 (2003).

\bibitem{taylor}
J. Parker et al.,
J. Phys. B {\bf 29}, L33 (1996);
D. Dundas et al.,
J. Phys. B {\bf 32}, L231 (1999);
J. Parker et al., 
Phys. Rev. Lett. {\bf 96}, 133001 (2006).

\bibitem{becker00kopold00}
A. Becker and F.~H.~M. Faisal, 
Phys. Rev. Lett. {\bf 84}, 3546 (2000);
R. Kopold et al.,
Phys. Rev. Lett. {\bf 85}, 3781 (2000).

\bibitem{alinged}
R. Grobe and J. H. Eberly, Phys. Rev. A {\bf 48}, 4664 (1993);
D. Bauer, Phys. Rev. A {\bf 56}, 3028 (1997);
D. G. Lappas and R. van Leeuwen, J. Phys. B {\bf 31}, L249 (1998);
W.-C. Liu et al.,
Phys. Rev. Lett. {\bf 83}, 520 (1999); 
S.~L. Haan et al.,
Phys. Rev. A {\bf 66}, 061402(R) (2002).

\bibitem{engel}
M. Lein, E. K. U. Gross, and V. Engel Phys. Rev. Lett. {\bf 85}, 4707 (2000).

\bibitem{becker}
C. Ruiz et al.,
Phys. Rev. Lett. {\bf 96}, 053001 (2006).  

\bibitem{ours}
K. Sacha and B. Eckhardt, Phys. Rev. A {\bf 63}, 043414 (2001);
B. Eckhardt and K. Sacha, Europhys. Lett. {\bf 56}, 651 (2001).

\bibitem{corkum93}
P.~B. Corkum, Phys. Rev. Lett. {\bf 71}, 1994 (1993);
K.C. Kulander, J. Cooper, and K.J. Schafer, 
Phys. Rev. A {\bf 51}, 561 (1995).

\bibitem{eckhardtunpubl}
B. Eckhardt and K. Sacha, arXiv:physics/0605195.

\bibitem{footnote} Actually, in the restricted model considered here, the saddle 
is a maximum in the potential, but we will keep the name 
since it corresponds to a saddle in the full 3d situation.

\bibitem{krausz} 
M. Nisoli et al.,
Opt. Lett. {\bf 22}, 522 (1997).

\end{thebibliography}
\end{document}